# Anisotropic Moiré Optical Transitions in Twisted Monolayer/bilayer Phosphorene Heterostructures

Shilong Zhao[1,2]†, Erqing Wang[1]†, Ebru Alime Üzer[3], Shuaifei Guo[4], Kenji Watanabe[5], Takashi Taniguchi[5], Tom Nilges[3], Yuanbo Zhang[4], Bilu Liu[1*], Xiaolong Zou[1*], and Feng Wang[1,2,6,7]*

1. Tsinghua-Berkeley Shenzhen Institute, Tsinghua University, Shenzhen, China
2. Department of Physics, University of California at Berkeley, Berkeley, CA, USA
3. Department of Chemistry, Technical University of Munich, Garching, Germany
4. State Key Laboratory of Surface Physics and Department of Physics, Fudan University, Shanghai, China
5. National Institute for Materials Science, Tsukuba, Japan
6. Material Science Division, Lawrence Berkeley National Laboratory, Berkeley, CA, USA
7. Kavli Energy NanoSciences Institute at University of California Berkeley and Lawrence Berkeley National Laboratory, Berkeley, CA, USA

† These authors contributed equally to this work

* Correspondence to: fengwang76@berkeley.edu, xlzou@sz.tsinghua.edu.cn, or bilu.liu@sz.tsinghua.edu.cn.

## Abstract

Moiré superlattices of van der Waals heterostructures provide a powerful new way to engineer the electronic structures of two-dimensional (2D) materials. Many novel quantum phenomena have emerged in different moiré heterostructures, such as correlated insulators[1,2], superconductors[3,4], and Chern insulators[5,6] in graphene systems and moiré excitons in transition metal dichalcogenide (TMDC) systems[7-10]. Twisted phosphorene offers another attractive system to explore moiré physics because phosphorene features an anisotropic rectangular lattice[11-14], different from the isotropic hexagonal lattice in graphene and TMDC. Here we report emerging anisotropic moiré optical transitions in twisted monolayer/bilayer phosphorene. The optical resonances in phosphorene moiré superlattice depend sensitively on the twist angle between the monolayer and bilayer. Surprisingly, even for a twist angle as large as 19° the moiré heterostructure exhibits optical resonances completely different from those in the constituent monolayer and bilayer phosphorene. The new moiré optical resonances exhibit strong linear polarization, with the principal axis lying close to but different from the optical axis of bilayer phosphorene. Our *ab initio* calculations reveal that the Γ-point direct bandgap and the rectangular lattice of phosphorene, unlike the K-point bandgap of hexagonal lattice in graphene and TMDC, give rise to the remarkably strong moiré physics in large-twist-angle phosphorene heterostructures. Our results highlight the exciting opportunities to explore moiré physics in phosphorene and other van der Waals heterostructures with different lattice configurations.

Stacked van der Waals heterostructures with a finite twist angle can generate a moiré superlattice, which is characterized by a periodic variation of the interlayer stacking order. Such moiré superlattices can dramatically modify the electronic band structures of two-dimensional (2D) van der Waals heterostructures and have given rise to many fascinating quantum phenomena. For example, correlated insulator states[1,2], superconductivity[3,4,15], magnetism[6,16], and Chern insulators[5,6] have been observed in "magic angle" twisted bilayer graphene and in ABC trilayer graphene/hexagonal boron nitride (hBN) moiré superlattices, while moiré excitons have been reported in small-twist-angle transition metal dichalcogenide (TMDC) heterostructures, e.g., $WS_2/WSe_2$[7], $MoSe_2/WS_2$[8], and $MoSe_2/WSe_2$ heterostructures[9,10]. All these materials (graphene, hBN, and TMDCs) belong to the 2D hexagonal structures with the electronic bandgap lying at the vertices (K and K' points) of the Brillouin zone[17-21], and prominent moiré superlattice effects have been observed in small-twist-angle ($\theta<2°$) heterostructures[1,3,7-10,22]. It will be highly desirable to explore new moiré heterostructures with different lattice configurations that may exhibit strong moiré physics even for large twist angles.

In this letter, we reported the first experimental study of rectangular moiré superlattices of twisted monolayer/bilayer phosphorene heterostructures. Phosphorene features a puckered honeycomb structure that forms an anisotropic rectangular unit cell[13,23,24], and it has a direct bandgap lying at the $\Gamma$ point in the first Brillouin zone[11-14]. In addition, few-layer phosphorene is known to exhibit unusually strong interlayer interactions, which leads to a dramatic change of the direct bandgap from 1.73 eV in monolayer phosphorene to 0.62 eV in tetralayer phosphorene[13,25]. Here we demonstrate the moiré potential completely changes the electronic band structure and gives rise to a new set of optical transitions in twisted monolayer/bilayer phosphorene heterostructure even for twist angles larger than 19°. This behavior is in striking contrast to other moiré systems, where prominent moiré effects exist only for twist angles smaller than a few degrees. The emerging optical resonances in the phosphorene moiré heterostructure are linearly polarized, and the polarization axis is closer to but different from the armchair direction of the bilayer phosphorene. Our *ab initio* density functional theory (DFT) calculations show that the remarkably large moiré superlattice potential at large twist angle originates from the Γ-point direct bandgap as well as the strong and stacking-dependent interlayer electron hybridization in twisted phosphorene heterostructures. The theory also reveals the extremely important role of the underlying electron Bloch wavefunction in the interlayer coupling, which results in a strongly hybridized conduction band but a negligibly coupled valence band in twisted phosphorene heterostructure. The calculated optical responses of the monolayer-bilayer moiré superlattice agree well with our experimental observations.

Figure 1a illustrates the configuration of a twisted monolayer/bilayer phosphorene heterostructure encapsulated between thin hBN layers. Few-layer phosphorene samples were first mechanically exfoliated onto the surface of polydimethylsiloxane (PDMS) thin films, and then transferred to 285-nm thick silicon dioxide/silicon ($SiO_2$/Si) substrates (see methods)[25,26]. The layer number of phosphorene was determined from the optical image contrast and verified by photoluminescence (PL) measurements. Thin hBN layers were also mechanically exfoliated from their bulk crystals. We then sequentially assembled the top hBN, monolayer phosphorene, bilayer phosphorene, and the bottom hBN using a dry-transfer method[27,28] (see Methods). The whole structure was then transferred onto a 90-nm thick $SiO_2$/Si substrate for further optical measurements. To minimize sample degeneration, the whole fabrication process was done inside a nitrogen-gas-filled glovebox with both moisture and oxygen levels lower than 0.1 ppm.

Figure 1b illustrates the moiré superlattice of the phosphorene heterostructure with a large twist angle of 19°. The gray dashed rectangle indicates the supercell of the moiré superlattice with four high symmetry points at A, B, C, and D (see Fig. S1 and S2 in Supplementary Information for the atom configurations of these high symmetry points). Fig. 1c shows the optical microscopy image of a representative device (D1), where the monolayer and bilayer regions are outlined with red and blue dashed line, respectively, and the twisted phosphorene heterostructure exists in the overlapping region outlined by white dashed lines (see Fig. S3 in Supplementary Information for details). We determined the twist angle between the monolayer and bilayer phosphorene through their anisotropic optical resonances. Figure 1d shows the polarization-dependent PL of the monolayer and bilayer phosphorene, from which we can determine a rotation of the principal axis of 19° ± 1° between the monolayer and bilayer. Two different devices (D2 and D3) with the monolayer-bilayer twist angle of 6° and 2° are shown in Fig. S4 (Supplementary Information). Fig. 1e display the PL spectra of the monolayer (red), bilayer (black), and heterostructure regions (blue) in the device D1, respectively. Surprisingly, we observe an emerging moiré optical transition at 0.83 eV in the twisted monolayer/bilayer phosphorene heterostructure, which is distinctly different from the monolayer resonance at 1.73 eV and the bilayer resonance at 1.10 eV. It shows that moiré superlattice has a dramatic effect on the optical properties of the phosphorene heterostructure even for twist angles as large as 19°.

Next we investigate optical resonances of the phosphorene heterostructures with different monolayer-bilayer twist angles using photoluminescence excitation (PLE) spectroscopy. Fig. 2a, 2b, 2c, and 2d show the 2D color plot of the PLE results for the twisted phosphorene heterostructure with the twist angle at 19°, 6°, 2°, and 0° (i.e. exfoliated trilayer), respectively. The color scale bar corresponds to the PL intensities. In addition, the right panels in the figure show the integrated PL intensity between 0.8 eV – 0.9 eV as a function of the excitation energy. The PLE spectra show that optical properties of the twisted monolayer/bilayer phosphorene depend strongly on the twist angles, and they exhibit optical transitions distinctly different from the monolayer and bilayer phosphorene. All the twisted phosphorene heterostructures show low energy PL emission close to 0.83 eV, but there is a small energy shift among heterostructures of different twist angles. In addition, the PL emission in large twist angle heterostructures (at 19° and 6°) is significantly broader. The high energy optical absorption in different heterostructures, however, are dramatically different: the 19° twisted heterostructure exhibit a broad absorption with no clear resonances, the 6° heterostructure has a weak absorption peak at 1.97 eV and a strong absorption peak at 2.64 eV, while the 2° heterostructure and trilayer phosphorene shows a single prominent absorption peak at 1.92 eV. The strongly twist-angle dependent optical transitions in monolayer/bilayer phosphorene heterostructures indicate that moiré potentials strongly modulate the electronic band structures of twisted phosphorenes, and the moiré superlattice effect remains strong even in large twist angle heterostructures, which is different from reported systems with isotropic hexagonal lattice structures.

We further investigated the polarization dependence of the PL emission in the 19° twisted phosphorene heterostructure (blue dots, Fig. 3a). The PL intensity shows a well-defined $\cos^2\theta$ pattern (blue line, Fig. 3a), which indicates a linearly polarized emission from the twisted phosphorene heterostructure. Interestingly, the polarization principal axis does not align with that of either the monolayer (red line) or the bilayer (black line). Instead, it is rotated by 4° (± 1°) from the bilayer polarization principal axis. This behavior further demonstrates the strong modulation effect of moiré superlattice on optical resonances of the 19° twisted phosphorene heterostructure.

To understand the unusual moiré superlattice effects on the electronic band structure and optical resonances of twisted phosphorene, we performed *ab initio* DFT calculations using the Vienna Ab initio Simulation Package (VASP[29], See Methods). We focused on the 19° heterostructure. Calculations on 6° and 2° twisted phosphorene are beyond our computation capability due to their much larger unit cell sizes. We first benchmarked our DFT calculations in pristine monolayer, bilayer, and trilayer phosphorene, and our results reproduce nicely the known optical transitions in these systems[13,30] (see Fig. S5 in Supplementary Information).

Figure 3b displays the calculated imaginary part of dielectric function (i.e., optical absorption spectrum) of 19° twisted phosphorene heterostructure. The lowest optical resonance in the twisted phosphorene heterostructures is at 0.82 eV, which agrees well with the observed PL emission peak (black arrow in Fig. 3b). In addition, the spectrum shows a rather broad absorption between 1.3 to 2.6 eV with very weak resonances, consistent with the mostly featureless absorption spectrum observed experimentally for 19° twisted heterostructure (black line Fig. 3b). We also evaluated the angle-dependent optical absorption in the 19° twisted phosphorene heterostructure. Our calculation shows that the optical absorption is highly polarized in the heterostructure (see Section S5 in Supplementary Information). The polarization principal axis of the heterostructure lies in between of the polarization principal axis of the constituent monolayer and bilayer, and is rotated from the bilayer axis by ~ 7°. This is in qualitative agreement with the observed linearly polarization of PL emission in the 19° twisted device.

Next we examine in detail the electronic structure of both the conduction and valence bands close to the Γ point in the twisted monolayer/bilayer phosphorene heterostructure, and compare it to the band structures of the monolayer, bilayer, and trilayer phosphorene. Figure 4a shows our calculation results for the monolayer (black solid line), bilayer (red solid line), trilayer (blue solid line) and 19° twisted heterostructure (orange solid line). The two horizontal dashed lines indicate the conduction band minimum (CBM) and the valence band maximum (VBM) of the bilayer phosphorene. It shows that the CBM decreases and the VBM increases progressively from monolayer to trilayer due to the interlayer coupling, as have been established in previous studies[13]. The band structure of the 19° twisted phosphorene heterostructure, however, exhibits a very different behavior. It has a VBM energy that is almost identical to that of the bilayer phosphorene, but a CBM energy much lower than those of both bilayer and trilayer phosphorene. The unusual electronic bands of the twisted phosphorene heterostructure are further illustrated in the partial charge density distribution of the CBM and VBM, as shown in Fig. 4c and 4d, respectively. Apparently, the conduction band is strongly hybridized between the monolayer and bilayer. Electrons at the CBM are mostly localized at the monolayer-bilayer interface, and the charge density exhibit strong periodic modulation commensurate with the moiré superlattice (Fig. 4c). The largest electron density is observed at the near-ABC-stacked regions in the moiré superlattice. Electrons at the VBM, however, are mostly localized in the bottom bilayer with negligible monolayer-bilayer couplings (Fig. 4d). Our calculations also reveal that although the 19° twisted heterostructure and trilayer phosphorene have similar bandgap, the underlying conduction and valence bands are completely different.

The very different CBM and VBM coupling between the monolayer and bilayer phosphorene in the twisted heterostructure originates from the different electron Bloch wavefunctions at the CBM and VBM, as illustrated schematically in Fig. 4e and 4f, respectively. The red and blue colors represent the wavefunctions from the top monolayer and bottom bilayer phosphorene, respectively. The sign of the electron wavefunctions is indicated by the filled or blank spindle. Here, we focus on the $p_z$ orbital components of the wavefunctions because of their dominant contribution to the wavefunctions at the CBM and VBM. (see section S6 or Fig. S6 in Supplementary Information). We find that the conduction band electron wavefunctions have the same sign at the monolayer-bilayer interface. Therefore, interlayer coupling at different atom sites adds up constructively, resulting in strong hybridization between the monolayer and bilayer phosphorene at the CBM. In particular, the electron wavefunctions have the largest overlap at the near-ABC-stacked regions, as indicated by the dashed rectangles in Fig. 4e. It leads to the highest CBM electron density in these regions in the moiré superlattice, as observed in Fig. 4c. In contrast, the electron wavefunctions at the VBM exhibit oscillating signs at the monolayer-bilayer interface (Fig. 4f). As a result, the electron coupling at different atom sites interferes destructively with each other, giving rise to an extremely weak interlayer coupling.

Consequently, the VBM electrons are mostly localized in the bilayer phosphorene with negligible hybridization with the monolayer (Fig. 4d).

In summary, our experimental observations reveal that moiré superlattices can strongly modulate the electronic band structures and optical transitions of the twisted phosphorene. The moiré superlattice effect remains strong even for a very large twist angle, highlighting new moiré physics that can emerge in van der Waals heterostructures with rectangular lattices.

## Methods

**Fabrication of twisted phosphorene structures**.

Bulk black phosphorus crystals were synthesized from red phosphorus with Sn and $SnI_4$ as transport agents using a chemical vapor transport method, and details can be found in previous work[31,32]. Few-layer phosphorene flakes were first exfoliated onto the surface of polydimethylsiloxane (PDMS) thin film (Gel-Pak) and then transferred onto 285-nm $SiO_2$/Si substrates. The layer numbers of few-layer phosphorenes were estimated by measuring their optical image contrast, as described in Ref. 13. Thin hBN flakes were also mechanically exfoliated onto 285-nm $SiO_2$/Si substrates. Polyethylene terephthalate (PET) stamps were employed to sequentially pick up the top hBN, monolayer phosphorene, bilayer phosphorene, bottom hBN at 60 °C. The final structures were then released onto a 90 nm $SiO_2$/Si substrate at 90 – 110 °C, followed by dissolving PET residues in dichloromethane for at least 12 hours[7]. To minimize the degradation of phosphorene flakes, the exfoliation, identification, and assembly were done within a day. Moreover, all fabrication processes were done inside a nitrogen-gas-filled glove box with both the oxygen and humidity levels less than 0.1 ppm. The whole structures were then put in a cryostat and pumped down to vacuum ($< 1\times10^{-6}$ mbar) for optical measurements.

**Optical measurements**

Samples were kept at the liquid nitrogen temperature for all optical measurements. PL measurements were conducted using a lab-build micro-PL setup in reflection geometry with a long working distance near-infrared objective (50x, N.A. 0.42) and the spectrometer equipped with both silicon and InGaAs detectors. PLE measurements were performed using a super-continuum laser (SC-Pro, YSL) as the excitation source. Tunable excitation light with the linewidth less than 0.5 nm was spectrally picked up by a grating and filtered out by suitable filters. The spot size of the focused light was ~2 μm. The result spectra were normalized to both integration time and incident power. For polarization-dependent PL measurements, linearly polarized light was set with a Glan-Thompson polarizer and rotated by a half-wave plate.

**First-principles calculations**

The supercell of twisted phosphorenes was constructed based on the coincidence site lattice theory[33]. The commensurate 20.04° twisted phosphorene heterostructure was used to model the electronic band structure and optical properties of the experimental 19° one.

We performed DFT calculations using the projector augmented wave method[34] and the plane-wave basis as implemented in the VASP[29]. The cutoff energy of the plane-wave basis was set to be 450 eV. The vacuum thickness was set to be 15 Å to avoid the interaction between adjacent layers. In addition, the supercells were fully optimized until the residual force per atom was less than 0.01 eVÅ$^{-1}$. For monolayer, bilayer, and trilayer phosphorene, Monkhorst-Pack $10 \times 8 \times 1$, $20 \times 16 \times 1$, and $50 \times 40 \times 1$ $k$-meshes were adopted to sample the first Brillouin zones, calculate electronic band structures, and obtain absorption spectra, respectively. For large-angle twisted phosphorene, the k-meshes were $5 \times 3 \times 1$. The non-local vdW density functional (vdW-DF)[35] with the optB88 exchange functional (optB88-vdW) [36,37] was employed to describe vdW interactions among layers. The modified Becke Johnson (mBJ)[38,39] exchange-correlation functional was adopted to get more accurate bandgap estimations. The charge gradient used in the mBJ calculations was extracted from the bulk black phosphorus, which gives $c=1.1592$.

## Acknowledgements

S.Z thanks the financial support from Tsinghua-Berkeley Shenzhen Institute (TBSI), Tsinghua University. Partial of device fabrication was supported by the National Key R&D Program of China (no. 2018YFA0307200), the National Natural Science Foundation of China (nos. 51722206 and 51920105002), and the Bureau of Industry and Information Technology of Shenzhen for the "2017 Graphene Manufacturing Innovation Center Project" (no. 201901171523). The theoretical part was financially supported by the National Key R&D Program of China (grant no. 2017YFB0701600), the National Natural Science Foundation of China (11974197), Shenzhen Basic Research Projects (no. JCYJ20170407155608882), and Guangdong Innovative and Entrepreneurial Research Team Program (grant no. 2017ZT07C341). S.G. and Y.Z. acknowledge financial support from National Key Research Program of China (grant nos. 2016YFA0300703 and 2018YFA0305600), NSF of China (grant nos. U1732274, 11527805 and 11421404), Shanghai Municipal Science and Technology Commission (grant no. 18JC1410300), and Strategic Priority Research Program of Chinese Academy of Sciences (grant no. XDB30000000). The growth of black phosphorus crystals was supported by the Deutsche Forschungsgemeinschaft (DFG, German Research Foundation) under Germany's Excellence Strategy e-conversion cluster EXC 2089/1-390776260. The growth of hexagonal boron nitride crystals was supported by the Elemental Strategy Initiative conducted by the MEXT, Japan and the CREST (JPMJCR15F3), JST.

We acknowledge Dr. Chaw-Keong Yong, M. Iqbal Bakti Utama, Dangqing Wang, Dr. Wenyu Zhao, Emma C. Regan, and Halleh B. Balch for their help on the optical measurements.

## Author contributions

F.W. conceived the project. F.W., X.Z., and B.L supervised the experimental and theoretical studies. S.Z. fabricated twisted phosphorene heterostructures and performed optical measurements. E.W. and X.Z. performed the *ab initio* DFT calculations. S.G. and Y. Z. helped with the fabrication of twisted phosphorene heterostructures. F.W., S.Z., and B.L. analyzed the experimental data. E.A.Ü. and T.N. grew black phosphorus crystals. K.W. and T.T. grew hexagonal boron nitride crystals. All authors discussed the results and wrote the manuscript.

## Competing interests

The authors declare no competing interests.

## Data and materials availability

The data that support the findings of this study are available from the corresponding author upon reasonable request.

## Corresponding author

Correspondence and Requests for materials should be addressed to:

Feng Wang, fengwang76@berkeley.edu, Xiaolong Zou, xlzou@sz.tsinghua.edu.cn, or Bilu Liu, bilu.liu@sz.tsinghua.edu.cn.

# Figures

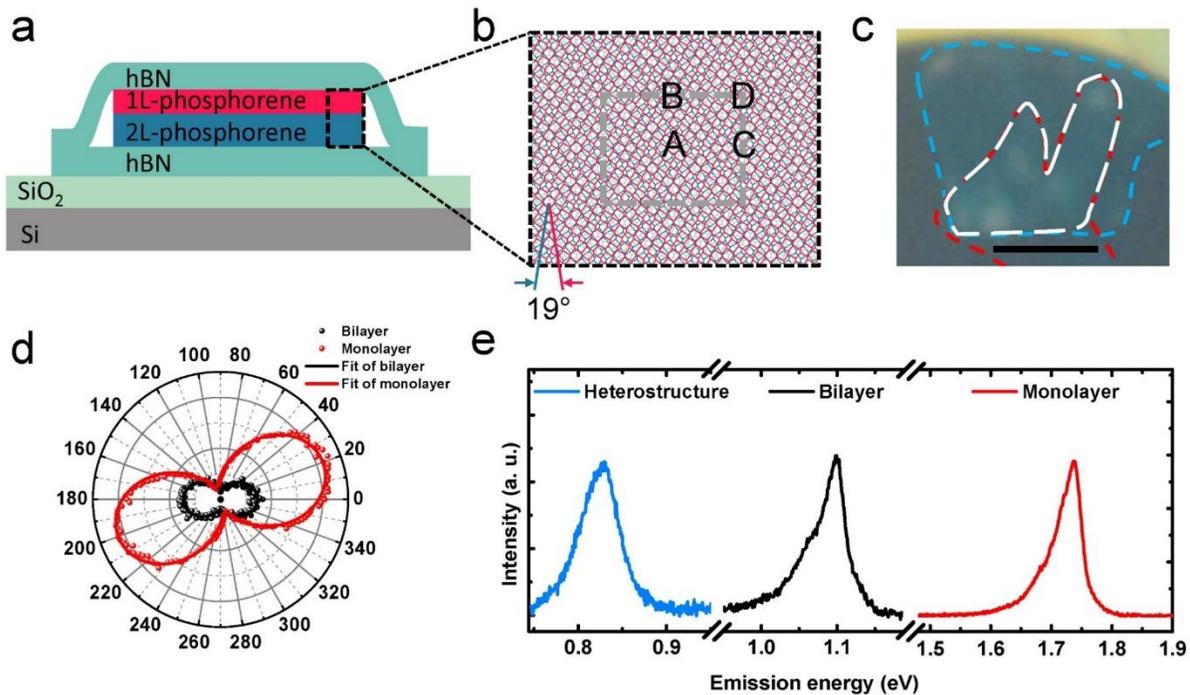

*Figure 1. Representative twisted monolayer/bilayer phosphorene heterostructures. (a), A side-view illustration of device's configuration. (b), A top-view schematic illustration of the twisted phosphorene heterostructure with a large angle of 19°. The top monolayer (red) is rotated by 19° respected to the bilayer phosphorene (light cyan for the middle layer and blue for the bottom layer). The gray dashed rectangle indicates the supercell that contains four high symmetric points, namely A, B, C, and D. (c), The optical image of device D1. The red, blue, and white dashed lines indicate the monolayer, bilayer, and the twisted phosphorene region, respectively. Scale bar, 5μm. (d), Angle-dependent PL emissions of the isolated monolayer and bilayer phosphorene, respectively. It shows a 19° twist angle between the top monolayer phosphorene and the bottom bilayer phosphorene. (e) PL spectra of the monolayer phosphorene (red), bilayer phosphorene (black), and the 19° twisted phosphorene heterostructure (blue). An emerging moiré optical transition at 0.83 eV is observed in the heterostructure, which is distinctly different from the*

*monolayer resonance at 1.73 eV and the bilayer resonance at 1.10 eV. It shows that moiré superlattice strongly modulates optical transition in phosphorene heterostructure even for twist angles as large as 19°.*

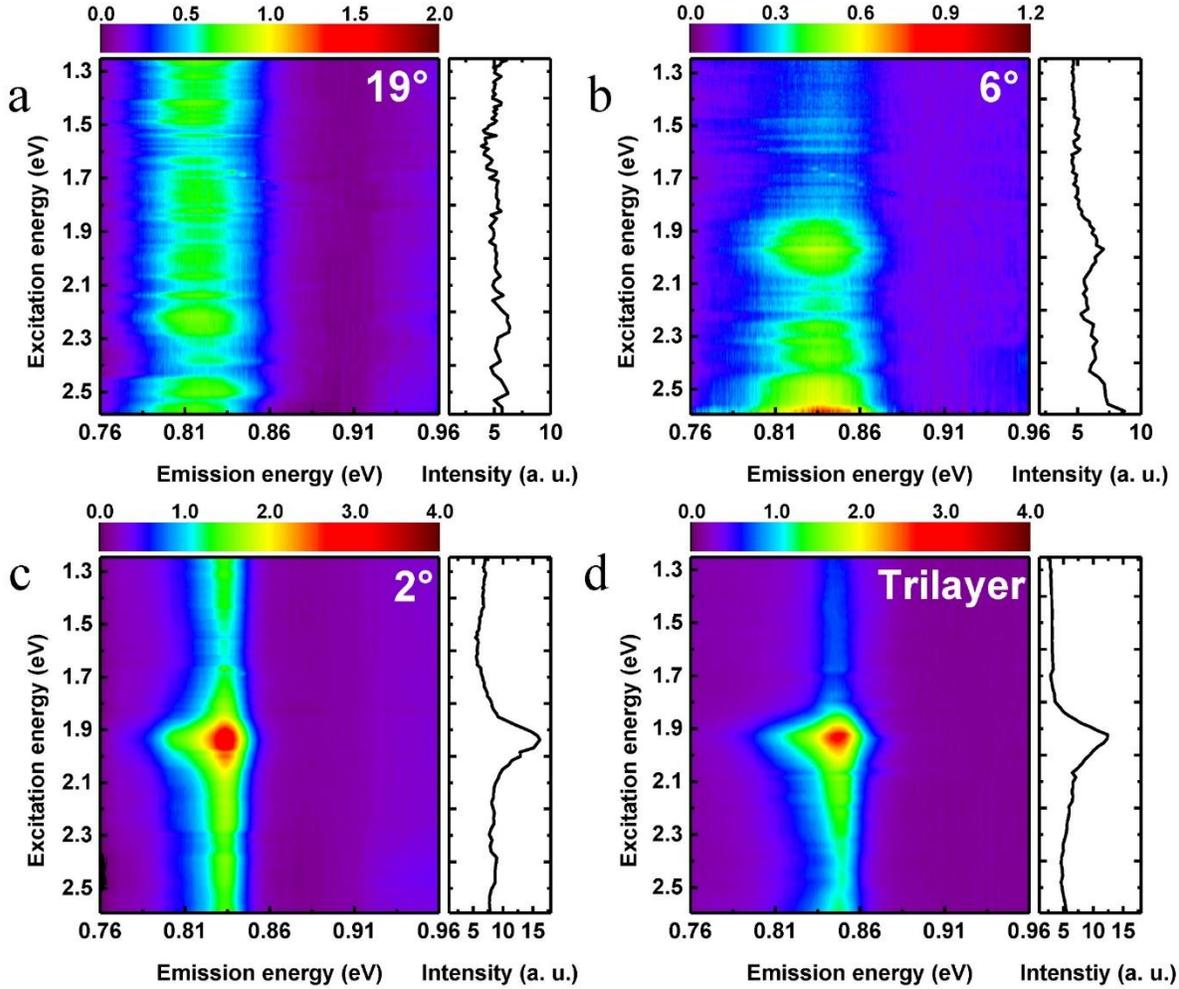

*Figure 2. Photoluminescence excitation spectra of twisted phosphorene heterostructures with the angle of 19° (a), 6° (b), 2° (c), and a trilayer phosphorene (d), respectively. The right panel in each figure shows the integrated PL emissions between the energy range of 0.8 – 0.9 eV. The top color scar indicates the PL emission intensities. All twisted phosphorene heterostructures show low energy PL emission close to 0.83 eV, but the high energy optical absorption spectra are completely different in different heterostructures: the 19° twisted heterostructure exhibit a broad*

*absorption with no clear resonances, the 6° heterostructure has a weak absorption peak at 1.97 eV and a strong absorption peak at 2.64 eV, while the 2° heterostructure and trilayer phosphorene shows a single prominent absorption peak at 1.92 eV.*

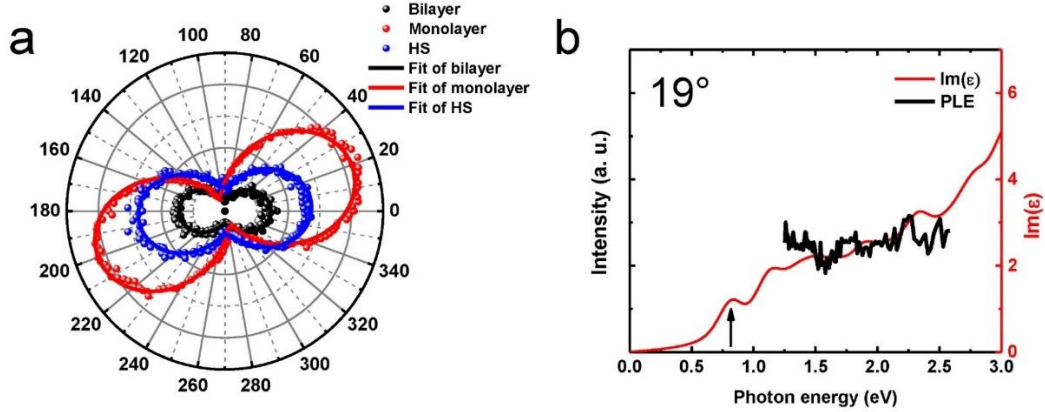

*Figure 3. Optical properties of the 19° twisted phosphorene heterostructure. (a), Polarization-dependent PL emissions (dots) and the corresponding fitting curves (solid lines) of the monolayer (red), bilayer (black), and the twisted phosphorene heterostructure (blue). The heterostructure shows a linearly polarized emission, and its principal axis is rotated by 4° from the bilayer polarization principal axis. (b), Comparison between the PLE spectra (black solid line) and the calculated imaginary parts of dielectric functions (red line) of the 19° twisted phosphorene heterostructure. The black arrow indicates the PL emission energy for the 19° twisted heterostructure.*

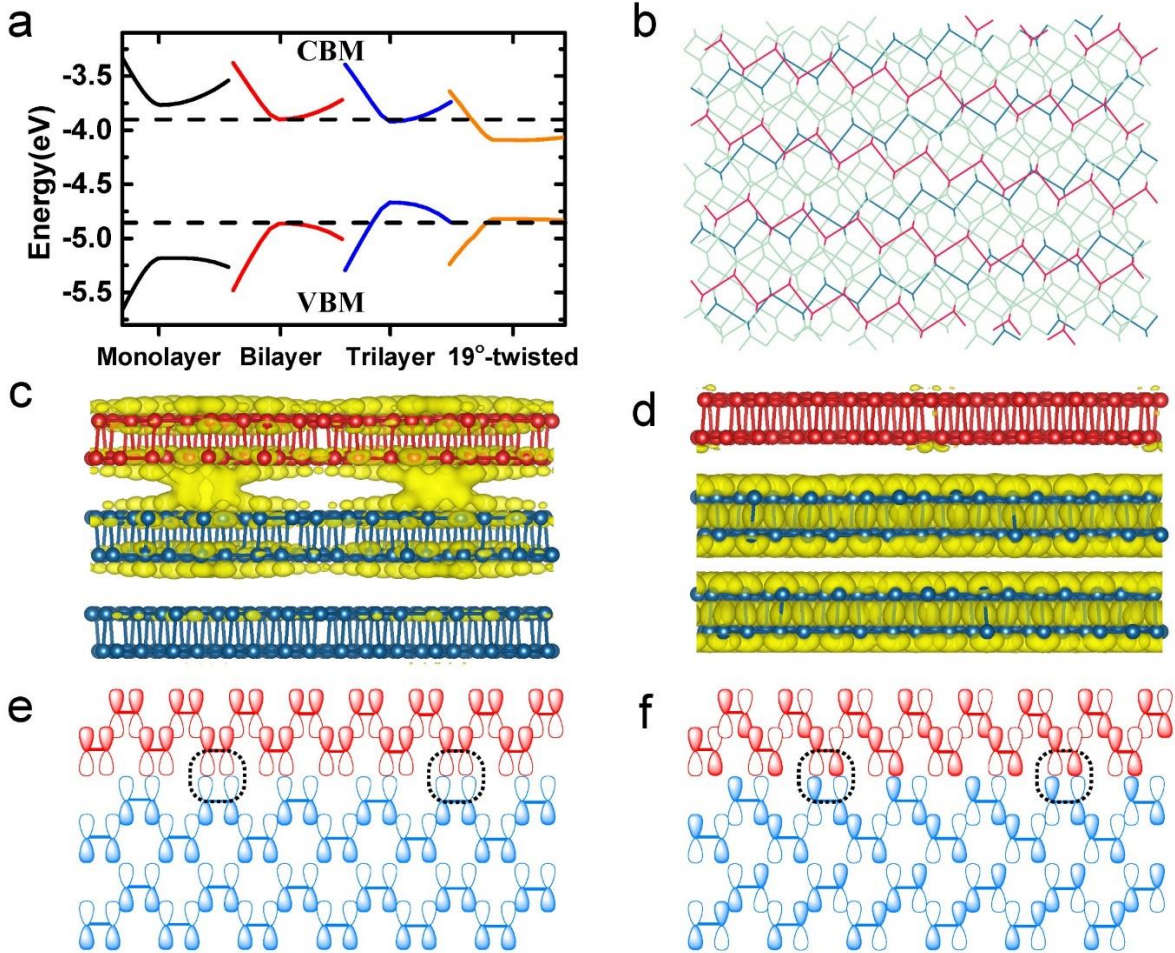

*Figure 4. (a), Band alignment of the monolayer (black solid line), bilayer (red solid line), trilayer (blue solid line), and 19° twisted phosphorene heterostructure (orange solid line) near Γ points. The two black dashed lines indicate the CBM and VBM of bilayer phosphorene. The vacuum energy is set to be at 0 eV. (b), The schematic top-view of the supercell of the 19° twisted phosphorene heterostructure. To clarify the stacking orders of atoms at the monolayer-bilayer interface, the bottom atoms from the top monolayer layer and the top atoms from the bottom layer are represented in red and blue, respectively. All other atoms are shown in light cyan. (c and d), The partial charge density distribution at the CBM (c) and VBM (d) of 19° twisted phosphorene heterostructure. The isosurface levels are set to be $1.5 \times 10^{-4}$ eV/Å. The wavefunctions of the monolayer (red) and bilayer phosphorene (blue) are strongly coupled at the CBM, whereas, they*

*are mainly localized in the bottom bilayer at the VBM. (e and f), Schematic illustration of the overlapping of wavefunctions at the CBM (e) and VBM (f) in 19° twisted phosphorene. The CBM electron wavefunctions have the same sign at the monolayer/bilayer interface (e), leading to constructive interference of interlayer coupling at different atom sites and strong hybridization between the monolayer and bilayer phosphorene. On the other hand, the VBM electron wavefunctions exhibit oscillating signs at the monolayer-bilayer interface (f), resulting in destructive interference of interlayer coupling at different atom sites and weak coupling of valence band states.*

# Supplementary Information for

**Anisotropic Moiré Optical Transitions in Twisted Monolayer/bilayer Phosphorene Heterostructures**

Shilong Zhao[1,2]†, Erqing Wang[1]†, Ebru Alime Üzer[3], Shuaifei Guo[4], Kenji Watanabe[5], Takashi Taniguchi[5], Tom Nilges[3], Yuanbo Zhang[4], Bilu Liu[1]*, Xiaolong Zou[1]*, and Feng Wang[1,2,6,7]*

† *These authors contributed equally to this work*

\* *Correspondence to: fengwang76@berkeley.edu, xlzou@sz.tsinghua.edu.cn, or bilu.liu@sz.tsinghua.edu.cn.*

S1. Atom configurations for high symmetry stacking orders

S2. Identification of twisted monolayer/bilayer phosphorene heterostructures

S3. Angle determination for twisted phosphorene heterostructures

S4. Calculated band structures and the imaginary part of dielectric function for few-layer phosphorene

S5. Determine the primary optical axis for the 19° twisted phosphorene heterostructure

S6. Orbital contributions at the CBM and VBM of the 19° twisted phosphorene heterostructure

S7. Partial charge distribution of monolayer and bilayer phosphorene

S1. Atom configurations for high symmetry stacking orders

The intrinsic stacking order of few-layer phosphorene is ABA stacking, as illustrated in Fig. S1a. The relative translation between the top monolayer and bottom bilayer phosphorene gives rise to different stacking orders, see Fig. S1 (b – f). In large-angle twisted phosphorene heterostructure, the stacking orders of the high-symmetric point A, B, C, and D in Fig. 1b (main text) are close to ABA (Fig. S1d), ABB (Fig. S1e), ABC (Fig.S1c), and ABD stacking (Fig.S1f), respectively. Therefore, the stacking orders of point A, B, C, and D in the 19° twisted heterostructure are referred as near-ABA (Fig. S2a), near-ABB (Fig. S2b), near-ABC (Fig. S2c), and near-ABD stacking (Fig. S2d), respectively.

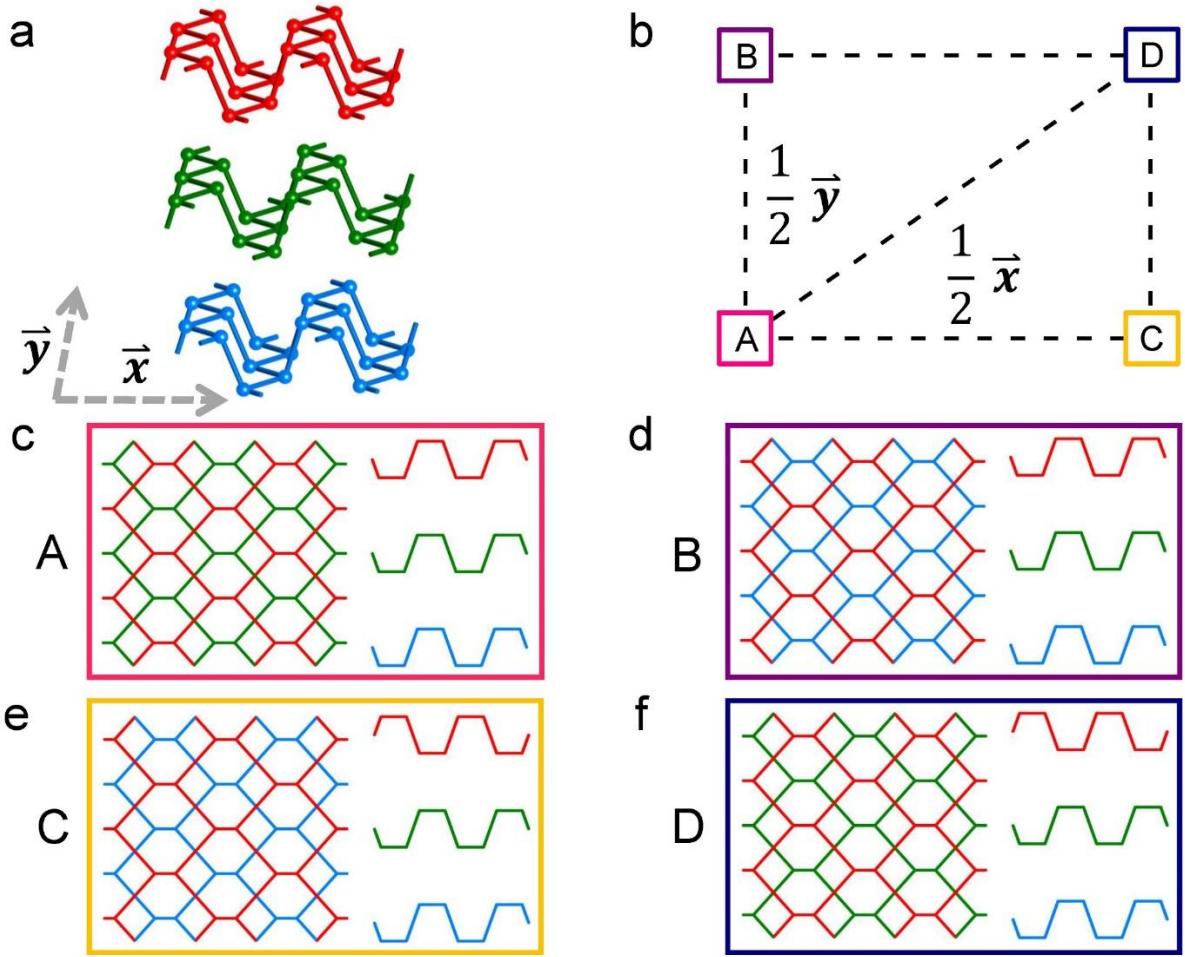

*Figure S1. Schematic illustration of the high symmetry stacking orders for monolayer/bilayer phosphorene heterostructures. (a), The schematic representation of the ABA stacking order. The blue, green and red indicate the 1$^{st}$, 2$^{nd}$, and 3$^{rd}$ layers in ABA stacking, respectively. In addition, armchair and zigzag direction are labeled as x- and y-direction, respectively. (b), Four high symmetry stacking orders in monolayer/bilayer phosphorene heterostructures. Generally, intrinsic trilayer phosphorene takes ABA staking order, as shown in (a). When the top monolayer is translated ½y along the y-direction, the result configuration is labeled as ABB stacking (purple, point B). Similarly, ABC stacking (orange, point C) forms if the translation is performed along x-direction with the length of ½x. In addition, the ABD stacking (dark blue, point D) forms when the top monolayer is shifted along the diagonal direction of the phosphorene lattice. The corresponding atom configurations of ABA, ABB, ABC, and ABD stacking orders are shown in (c), (d), (e), and (f), respectively. The red, green and blue colors in (c – f) indicate the top, middle, and*

*bottom layer of phosphorene, respectively. Moreover, the left and right panel in (c – f) represent the top-view and side-view of these structures.*

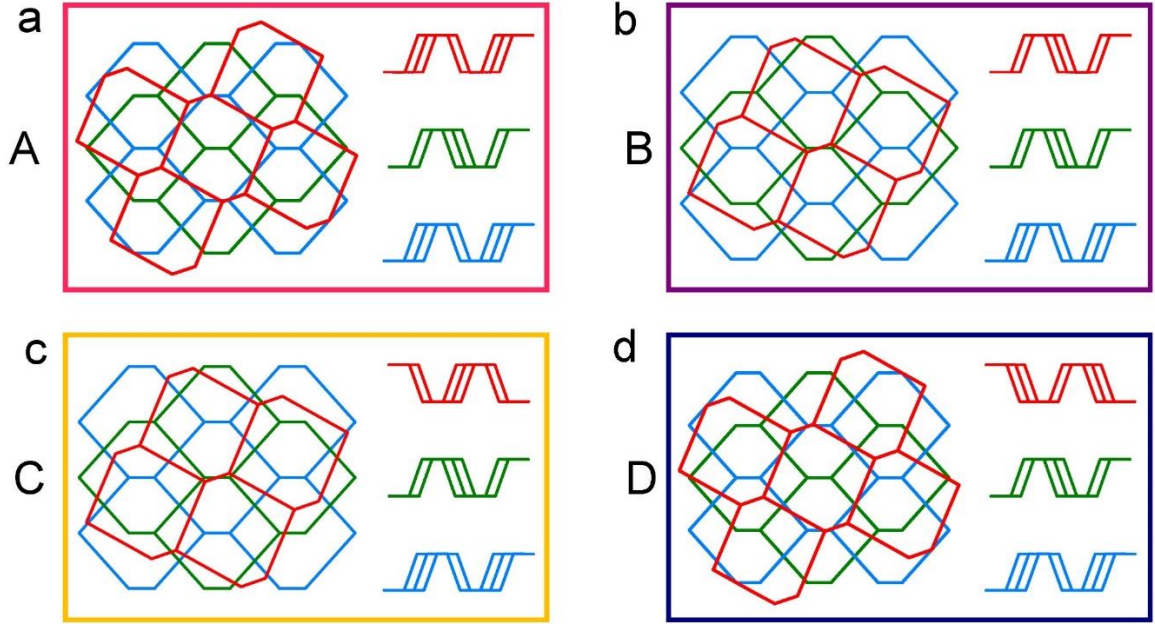

*Figure S2. Schematic illustration of the atom configurations for four high-symmetric points in the 19° twisted monolayer/bilayer phosphorene heterostructure. (a-d), The atom configurations (left, top-view, right, side-view) of the near-ABA (a), near-ABB (b), near-ABC (c), and near-ABD (d) stacked regions that correspond to the point A, B, C, and D in the main text, respectively. The red, green, and blue color in (a – d) represent the top, middle, and bottom layer of the 19° twisted phosphorene heterostructure, respectively.*

S2. Identification of twisted monolayer/bilayer phosphorene heterostructures

Fig. S3a, S3b, and S3c show the optical images of monolayer, bilayer phosphorene (before stacking), and their twisted heterostructure, respectively. The red and blue dashed lines represent the monolayer and bilayer phosphorene, respectively. The monolayer phosphorene broke during transfer, giving rise to the irregular shape shown in Fig. S3c and S3d. We further identified different regions by mapping their PL emissions at 77 K. As an example, Fig. S3e shows the bilayer

region in device D1, which matches well with the optical images shown in Fig. S3d. Markedly, strong PL emission was observed from the twisted heterostructure region, as shown in Fig. S3f. Moreover, the strongly quenched area of the bilayer phosphorene matched well with the PL emission area of the twisted heterostructure, indicating the formation of heterostructures.

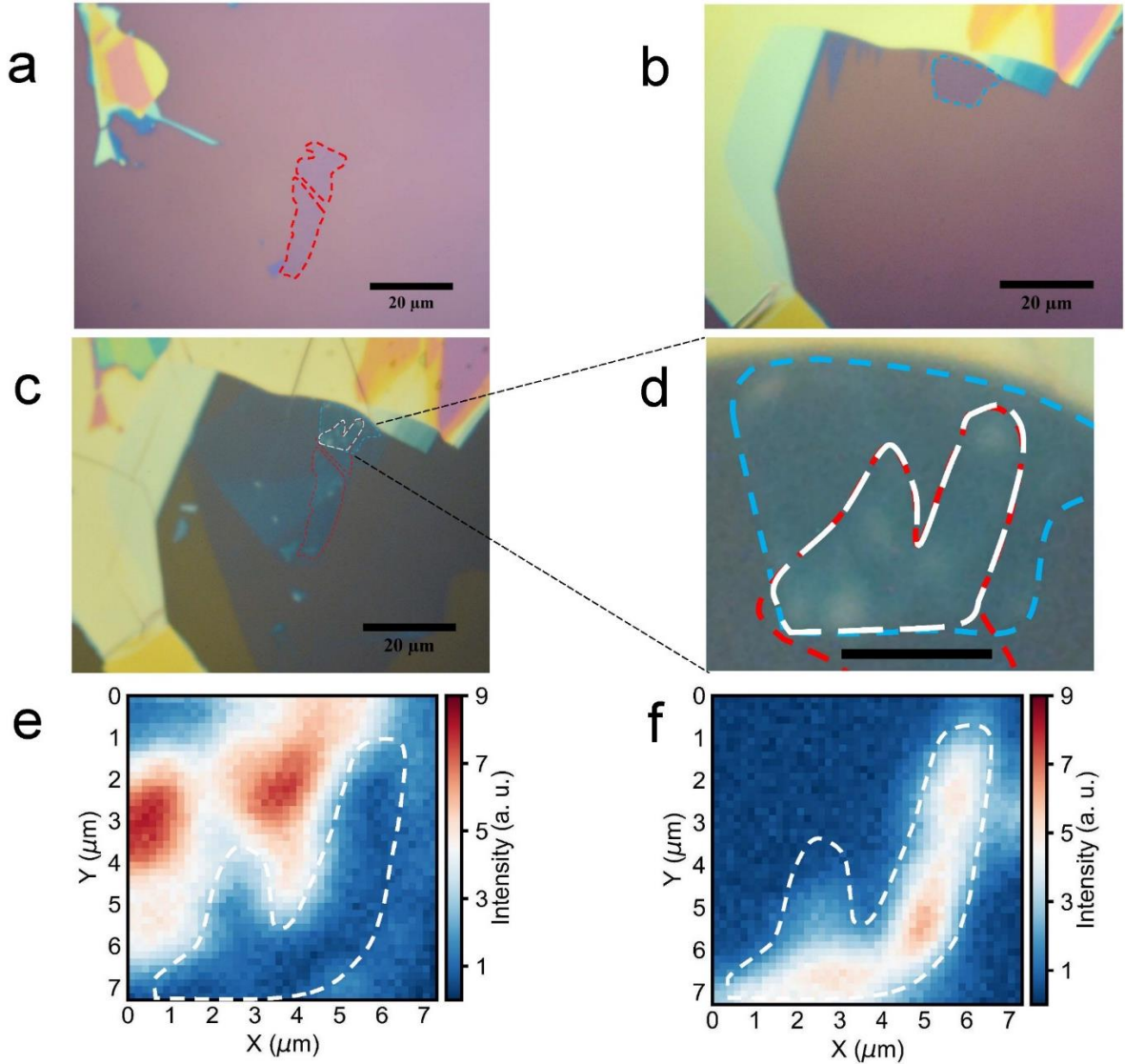

*Figure S3. The optical images of the monolayer (a), bilayer phosphorene (b) before stacking, and their twisted structure (c). Red and blue dashed lines indicate the monolayer (a) and bilayer phosphorene flakes (b), respectively. The monolayer phosphorene broke during transfer, giving rise to the irregular shape. (d), the zoom-in image of (c) with the twisted phosphorene*

*heterostructure highlighted by the white dashed line. (e and f) the PL mapping results using PL emission peaks of bilayer (e) and twisted phosphorene heterostructure (f). The PL quenching region in (e) matches well with the emission region of the twisted phosphorene heterostructures, as shown in white dashed lines.*

S3. Angle determination for twisted phosphorene heterostructures

Two more devices (D2 and D3) were fabricated using the same method described in the main text. Fig. S4a and S4b show the optical image of D2 and D3, respectively. The white dashed and solid lines indicated the monolayer and bilayer phosphorene regions, respectively.

We fit the experimental angle-dependent PL emission spectra with the function:

$$I = I_0 * cos^2(\theta + \varphi) + C \qquad (1),$$

where I represents the PL emission intensity, $I_0$ and C are constants to be fitted, $\theta$ and $\varphi$ indicate the incident angle of polarized light and the phase angle, respectively. Based on this method, the relative angle difference between monolayer and bilayer in device D1, D2 and D3 are 19° ± 1.1°, 5.9° ± 0.8°, and 2.2° ± 0.9°, respectively, see Fig. 1f in the main text, Fig. S4c and S4d.

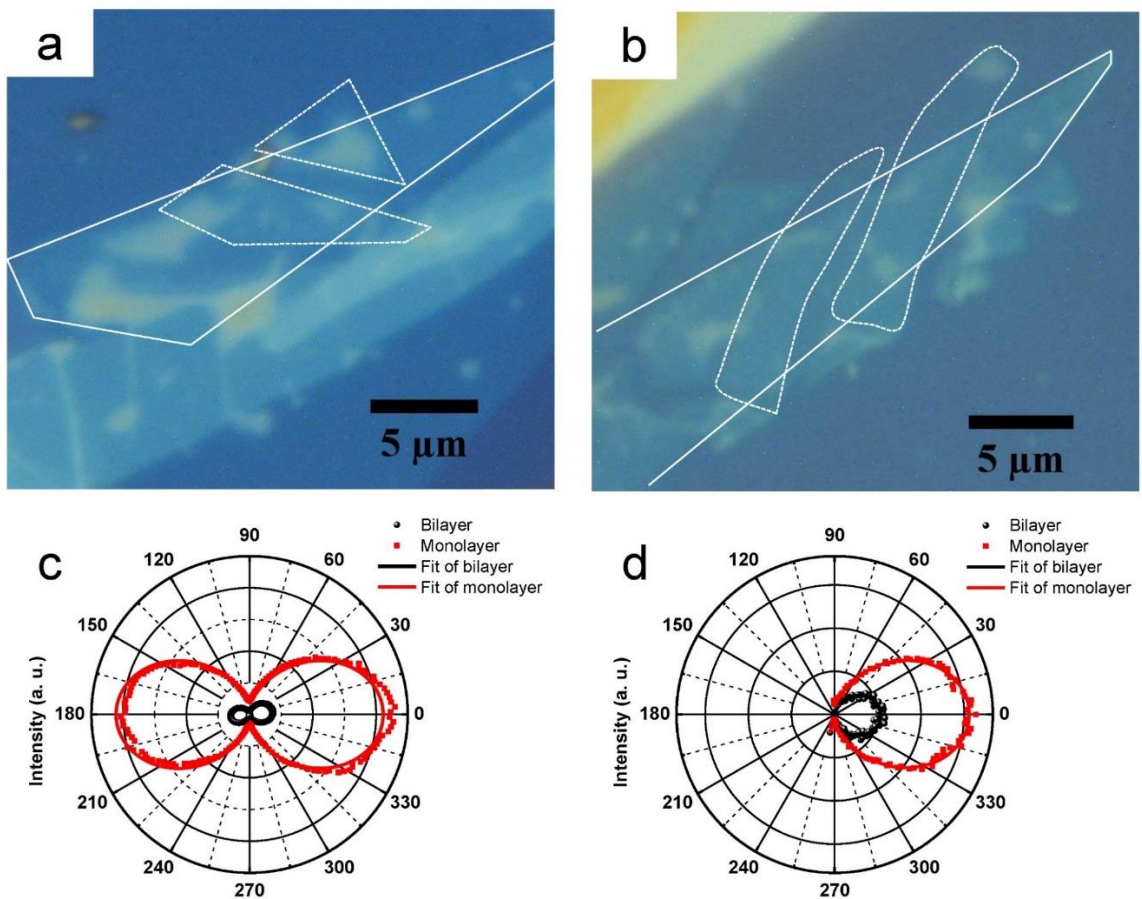

*Figure S4. Characterization of D2 and D3. (a and b), the optical images of D2 (a) and D3 (b). The white dashed and solid lines indicate the monolayer and bilayer phosphorene, respectively. (c and d), angle-dependent PL emission of the monolayer (red square) and bilayer phosphorene (black sphere) in D2 (a) and D3 (b). The red and black solid lines represent the fitting curves of angle-dependent PL emission for monolayer and bilayer phosphorene, respectively.*

S4. Calculated band structures and the imaginary parts of dielectric functions for few-layer phosphorene.

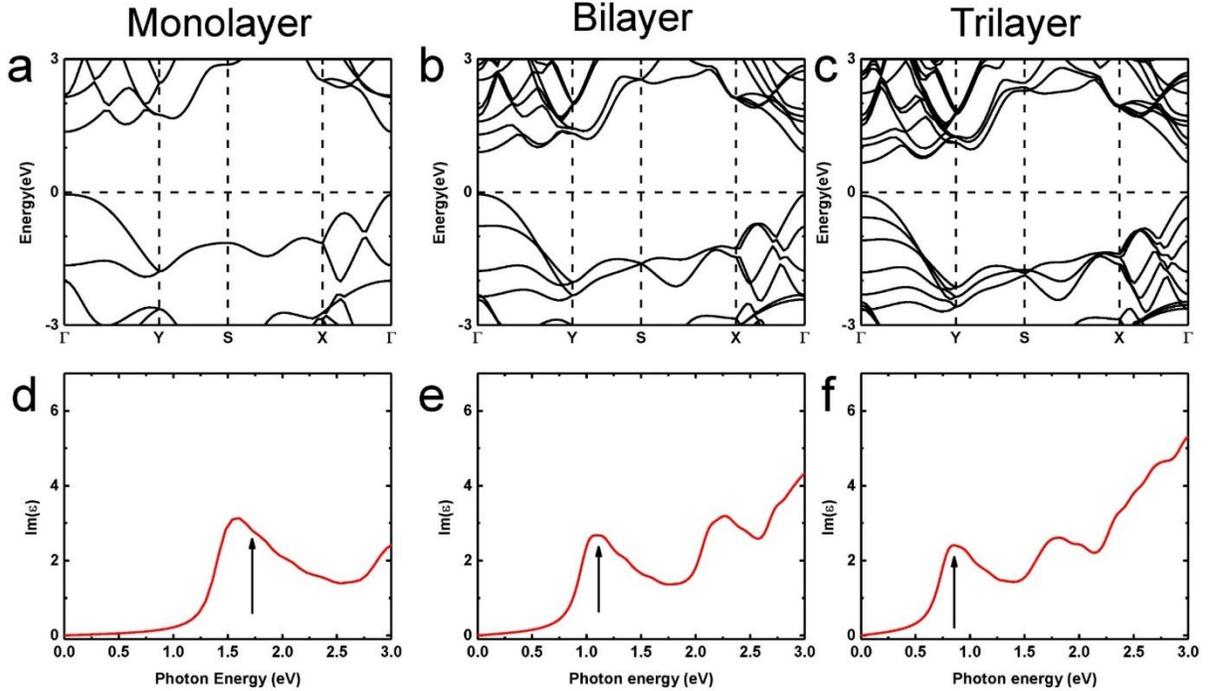

*Figure S5. The calculated band structures (a – c) and the imaginary part of dielectric functions (d – f) for monolayer (a and d), bilayer (b and e), and trilayer phosphorene (c and f) using mBJ functional. The black arrow in (d – f) indicates the PL emission energy of monolayer (d), bilayer (e) and trilayer phosphorene (f). In addition, the above bandgap transitions can be identified for both bilayer (e) and trilayer phosphorene (f)[1,2], respectively.*

S5. Determine the primary optical axis for the 19° twisted phosphorene heterostructure.

A 20.04° commensurate twisted phosphorene heterostructure was used to model our experimental 19° one. The x-direction of the 20.04° twisted phosphorene heterostructure laid in the middle of the armchair direction of its constituent monolayer (-10.02°) and that of bilayer phosphorene (10.02°). The calculated dielectric functions for 20.04° twisted heterostructure had non-zero components along xy-direction (blue lines, Fig. S6), which indicated the primary optical axis of the twisted heterostructure was away from its x-direction. The calculated x-, y-, and xy-component of the imaginary part of the dielectric functions read 1.8456, 0.5230, and 0.0741 at the energy of 2.33 eV. The diagonalization of the absorption matrix (M) using the rotation matrix (R) with an angle of φ gave rise to the angle difference (with value of -φ) between the primary optical axis and the x-direction of the twisted phosphorene heterostructure, where matrix M and R were constructed by equation (2) and (3).

$$M = \begin{bmatrix} xx & xy \\ xy & yy \end{bmatrix} = \begin{bmatrix} 1.8456 & 0.0741 \\ 0.0741 & 0.5230 \end{bmatrix} \quad (2),$$

$$R = \begin{bmatrix} \cos(\varphi) & -\sin(\varphi) \\ \sin(\varphi) & \cos(\varphi) \end{bmatrix} \quad (3),$$

The diagonalization was done using equation 4,

$$R^{-1}MR = \Lambda \quad (4),$$

where $\Lambda$ indicated a diagonal matrix. The calculation then gave φ ≈ -3.1°, which indicates that the primary optical axis of the twisted phosphorene heterostructure locates close to that of its constituent bilayer phosphorene with a ~7° angle difference.

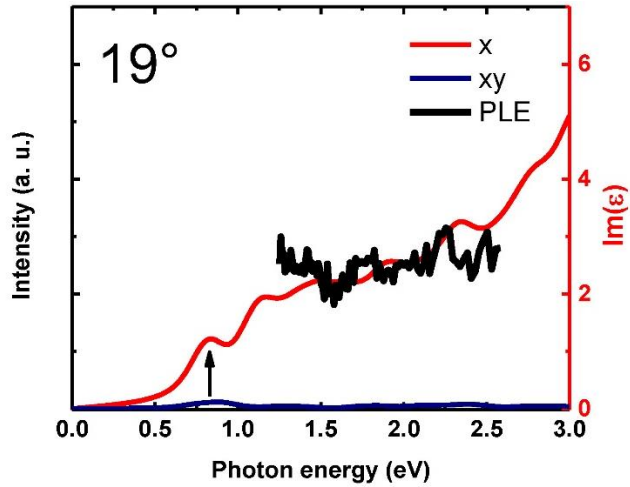

*Figure S6. The x- (red) and xy- (blue) component of the calculated dielectric function of the twisted heterostructure. The non-zero xy-component can be easily identified.*

S6. Orbital contributions at the CBM and VBM of the 19° twisted phosphorene heterostructure

P$_z$ orbitals contribute mainly to the wavefunctions at the CBM and VBM of the 19° twisted phosphorene structure, as shown in Table 1.

*Table 1. Orbital contributions at the CBM and VBM of the 19° twisted heterostructure*

|  | s | p$_x$ | p$_y$ | **p$_z$** | total |
|---|---|---|---|---|---|
| *CBM* | 0.058 | 0.004 | 0.072 | **0.207** | 0.341 |
| *VBM* | 0.058 | 0.001 | 0.013 | **0.44** | 0.513 |

S7. Partial charge density distribution of monolayer and bilayer phosphorene

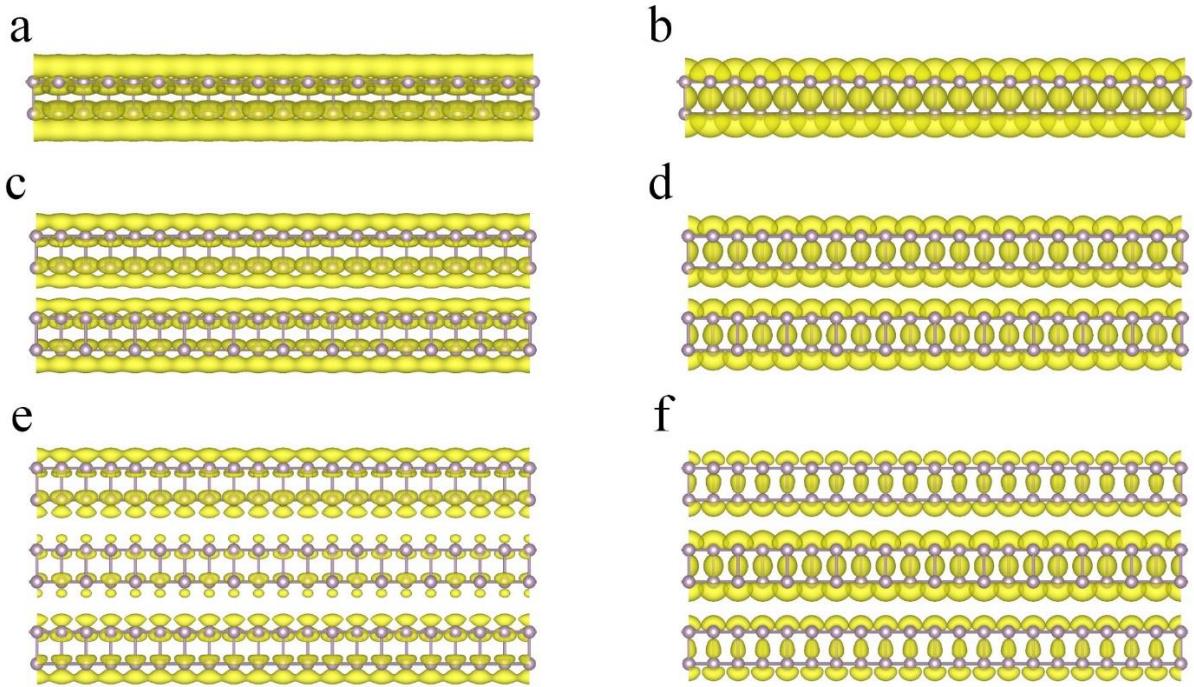

*Figure S7. The partial charge density distribution in the CBM (a and c) and VBM (b and d) of monolayer (a and b), bilayer (c and d), and trilayer phosphorenes (e and f), respectively. All the isosurface levels set to be 3.5 × 10$^{-3}$ eV/Å.*

# Supplementary References